\documentclass[twocolumn,aps,prl,showpacs,preprintnumbers,
superscriptaddress,floatfix]{revtex4}

\usepackage{graphicx}
\usepackage{dcolumn}
\usepackage{bm}

\begin{document}

%
%
\newcommand{\snn}{\sqrt{s_{NN}}}
\newcommand{\pbarp}{p(\overline{p})+p}
\newcommand{\zvtx}{z_{vtx}}
\newcommand{\npart}{N_{part}}
\newcommand{\ncoll}{N_{coll}}
\newcommand{\avenp}{\langle\npart\rangle}
\newcommand{\avenc}{\langle\ncoll\rangle}
\newcommand{\half}{\frac{1}{2}}
\newcommand{\halfnp}{(\half\avenp)}
\newcommand{\nch}{N_{ch}}
\newcommand{\etazero}{\eta = 0}
\newcommand{\etaone}{|\eta| < 1}
\newcommand{\dndeta}{d\nch/d\eta}
\newcommand{\dndetazero}{\dndeta|_{\etazero}}
\newcommand{\dndetaone}{\dndeta|_{\etaone}}
\newcommand{\dndetanp}{\dndeta / \halfnp}
\newcommand{\dndetaonp}{\dndeta / \np}
\newcommand{\dndetazeronp}{\dndetazero / \halfnp}
\newcommand{\dndetaonenp}{\dndetaone / \halfnp}
\newcommand{\Ratio}{R$_{200/19.6}$ = 2.03$\pm$0.02(stat)$\pm$0.05(syst)}


\title{Collision geometry scaling of $Au+Au$ pseudorapidity density \\ 
from $\snn$ = 19.6 to 200 GeV}

\author{B.B.Back}
\affiliation{Physics Division, Argonne National Laboratory, Argonne, IL 60439-4843, USA}
\author{M.D.Baker}
\affiliation{Chemistry and C-A Departments, Brookhaven National Laboratory, Upton, NY 11973-5000, USA}
\author{M.Ballintijn}
\affiliation{Laboratory for Nuclear Science, Massachusetts Institute of Technology, Cambridge, MA 02139-4307, USA}
\author{D.S.Barton}
\affiliation{Chemistry and C-A Departments, Brookhaven National Laboratory, Upton, NY 11973-5000, USA}
\author{R.R.Betts}
\affiliation{Department of Physics, University of Illinois at Chicago, Chicago, IL 60607-7059, USA}
\author{A.A.Bickley}
\affiliation{Department of Chemistry, University of Maryland, College Park, MD 20742, USA}
\author{R.Bindel}
\affiliation{Department of Chemistry, University of Maryland, College Park, MD 20742, USA}
\author{A.Budzanowski}
\affiliation{Institute of Nuclear Physics PAN, Krak\'{o}w, Poland}
\author{W.Busza}
\affiliation{Laboratory for Nuclear Science, Massachusetts Institute of Technology, Cambridge, MA 02139-4307, USA}
\author{A.Carroll}
\affiliation{Chemistry and C-A Departments, Brookhaven National Laboratory, Upton, NY 11973-5000, USA}
\author{M.P.Decowski}
\affiliation{Laboratory for Nuclear Science, Massachusetts Institute of Technology, Cambridge, MA 02139-4307, USA}
\author{E.Garc\'{\i}a}
\affiliation{Department of Physics, University of Illinois at Chicago, Chicago, IL 60607-7059, USA}
\author{N.George}
\affiliation{Physics Division, Argonne National Laboratory, Argonne, IL 60439-4843, USA}
\affiliation{Chemistry and C-A Departments, Brookhaven National Laboratory, Upton, NY 11973-5000, USA}
\author{K.Gulbrandsen}
\affiliation{Laboratory for Nuclear Science, Massachusetts Institute of Technology, Cambridge, MA 02139-4307, USA}
\author{S.Gushue}
\affiliation{Chemistry and C-A Departments, Brookhaven National Laboratory, Upton, NY 11973-5000, USA}
\author{C.Halliwell}
\affiliation{Department of Physics, University of Illinois at Chicago, Chicago, IL 60607-7059, USA}
\author{J.Hamblen}
\affiliation{Department of Physics and Astronomy, University of Rochester, Rochester, NY 14627, USA}
\author{G.A.Heintzelman}
\affiliation{Chemistry and C-A Departments, Brookhaven National Laboratory, Upton, NY 11973-5000, USA}
\author{C.Henderson}
\affiliation{Laboratory for Nuclear Science, Massachusetts Institute of Technology, Cambridge, MA 02139-4307, USA}
\author{D.J.Hofman}
\affiliation{Department of Physics, University of Illinois at Chicago, Chicago, IL 60607-7059, USA}
\author{R.S.Hollis}
\affiliation{Department of Physics, University of Illinois at Chicago, Chicago, IL 60607-7059, USA}
\author{R.Ho\l y\'{n}ski}
\affiliation{Institute of Nuclear Physics PAN, Krak\'{o}w, Poland}
\author{B.Holzman}
\affiliation{Chemistry and C-A Departments, Brookhaven National Laboratory, Upton, NY 11973-5000, USA}
\author{A.Iordanova}
\affiliation{Department of Physics, University of Illinois at Chicago, Chicago, IL 60607-7059, USA}
\author{E.Johnson}
\affiliation{Department of Physics and Astronomy, University of Rochester, Rochester, NY 14627, USA}
\author{J.L.Kane}
\affiliation{Laboratory for Nuclear Science, Massachusetts Institute of Technology, Cambridge, MA 02139-4307, USA}
\author{J.Katzy}
\affiliation{Laboratory for Nuclear Science, Massachusetts Institute of Technology, Cambridge, MA 02139-4307, USA}
\affiliation{Department of Physics, University of Illinois at Chicago, Chicago, IL 60607-7059, USA}
\author{N.Khan}
\affiliation{Department of Physics and Astronomy, University of Rochester, Rochester, NY 14627, USA}
\author{W.Kucewicz}
\affiliation{Department of Physics, University of Illinois at Chicago, Chicago, IL 60607-7059, USA}
\author{P.Kulinich}
\affiliation{Laboratory for Nuclear Science, Massachusetts Institute of Technology, Cambridge, MA 02139-4307, USA}
\author{C.M.Kuo}
\affiliation{Department of Physics, National Central University, Chung-Li, Taiwan}
\author{W.T.Lin}
\affiliation{Department of Physics, National Central University, Chung-Li, Taiwan}
\author{S.Manly}
\affiliation{Department of Physics and Astronomy, University of Rochester, Rochester, NY 14627, USA}
\author{D.McLeod}
\affiliation{Department of Physics, University of Illinois at Chicago, Chicago, IL 60607-7059, USA}
\author{A.C.Mignerey}
\affiliation{Department of Chemistry, University of Maryland, College Park, MD 20742, USA}
\author{R.Nouicer}
\affiliation{Department of Physics, University of Illinois at Chicago, Chicago, IL 60607-7059, USA}
\author{A.Olszewski}
\affiliation{Institute of Nuclear Physics PAN, Krak\'{o}w, Poland}
\author{R.Pak}
\affiliation{Chemistry and C-A Departments, Brookhaven National Laboratory, Upton, NY 11973-5000, USA}
\author{I.C.Park}
\affiliation{Department of Physics and Astronomy, University of Rochester, Rochester, NY 14627, USA}
\author{H.Pernegger}
\affiliation{Laboratory for Nuclear Science, Massachusetts Institute of Technology, Cambridge, MA 02139-4307, USA}
\author{C.Reed}
\affiliation{Laboratory for Nuclear Science, Massachusetts Institute of Technology, Cambridge, MA 02139-4307, USA}
\author{L.P.Remsberg}
\affiliation{Chemistry and C-A Departments, Brookhaven National Laboratory, Upton, NY 11973-5000, USA}
\author{M.Reuter}
\affiliation{Department of Physics, University of Illinois at Chicago, Chicago, IL 60607-7059, USA}
\author{C.Roland}
\affiliation{Laboratory for Nuclear Science, Massachusetts Institute of Technology, Cambridge, MA 02139-4307, USA}
\author{G.Roland}
\affiliation{Laboratory for Nuclear Science, Massachusetts Institute of Technology, Cambridge, MA 02139-4307, USA}
\author{L.Rosenberg}
\affiliation{Laboratory for Nuclear Science, Massachusetts Institute of Technology, Cambridge, MA 02139-4307, USA}
\author{J.Sagerer}
\affiliation{Department of Physics, University of Illinois at Chicago, Chicago, IL 60607-7059, USA}
\author{P.Sarin}
\affiliation{Laboratory for Nuclear Science, Massachusetts Institute of Technology, Cambridge, MA 02139-4307, USA}
\author{P.Sawicki}
\affiliation{Institute of Nuclear Physics PAN, Krak\'{o}w, Poland}
\author{W.Skulski}
\affiliation{Department of Physics and Astronomy, University of Rochester, Rochester, NY 14627, USA}
\author{P.Steinberg}
\affiliation{Chemistry and C-A Departments, Brookhaven National Laboratory, Upton, NY 11973-5000, USA}
\author{G.S.F.Stephans}
\affiliation{Laboratory for Nuclear Science, Massachusetts Institute of Technology, Cambridge, MA 02139-4307, USA}
\author{A.Sukhanov}
\affiliation{Chemistry and C-A Departments, Brookhaven National Laboratory, Upton, NY 11973-5000, USA}
\author{M.B.Tonjes}
\affiliation{Department of Chemistry, University of Maryland, College Park, MD 20742, USA}
\author{J.-L.Tang}
\affiliation{Department of Physics, National Central University, Chung-Li, Taiwan}
\author{A.Trzupek}
\affiliation{Institute of Nuclear Physics PAN, Krak\'{o}w, Poland}
\author{C.Vale}
\affiliation{Laboratory for Nuclear Science, Massachusetts Institute of Technology, Cambridge, MA 02139-4307, USA}
\author{G.J.van~Nieuwenhuizen}
\affiliation{Laboratory for Nuclear Science, Massachusetts Institute of Technology, Cambridge, MA 02139-4307, USA}
\author{R.Verdier}
\affiliation{Laboratory for Nuclear Science, Massachusetts Institute of Technology, Cambridge, MA 02139-4307, USA}
\author{F.L.H.Wolfs}
\affiliation{Department of Physics and Astronomy, University of Rochester, Rochester, NY 14627, USA}
\author{B.Wosiek}
\affiliation{Institute of Nuclear Physics PAN, Krak\'{o}w, Poland}
\author{K.Wo\'{z}niak}
\affiliation{Institute of Nuclear Physics PAN, Krak\'{o}w, Poland}
\author{A.H.Wuosmaa}
\affiliation{Physics Division, Argonne National Laboratory, Argonne, IL 60439-4843, USA}
\author{B.Wys\l ouch}
\affiliation{Laboratory for Nuclear Science, Massachusetts Institute of Technology, Cambridge, MA 02139-4307, USA}

\collaboration{PHOBOS Collaboration}
\noaffiliation

\date{\today}

\begin{abstract}
The centrality dependence of the midrapidity charged particle 
multiplicity in $Au+Au$ collisions at $\snn$ = 19.6 and 
200 GeV is presented.  Within a simple model, the fraction of 
hard (scaling with number of binary collisions) 
to soft (scaling with number of participant pairs) 
interactions is consistent with a value of 
$x$ = 0.13$\pm$0.01(stat)$\pm$0.05(syst) at both energies.
The experimental results at both energies, scaled by inelastic
$\pbarp$ collision data, agree within systematic errors.
The ratio of the data was found not to depend on centrality
over the studied range and yields 
a simple linear scale factor of \Ratio.  
\end{abstract}

\maketitle

We have studied the centrality dependence of the charged particle
multiplicity at midrapidity for $Au+Au$ collisions
at nucleon-nucleon center of mass energies $\snn =$ 19.6 GeV and 
200 GeV, using the PHOBOS detector.  The data were taken during 
the $Au+Au$ run in 2001-2002 at the Relativistic Heavy Ion 
Collider (RHIC) in Brookhaven National Laboratory.  
Data at both energies have allowed the extraction of 
results with the same detector, which covers a factor of ten in 
collision energy, from slightly above the highest energy of  
the CERN SPS fixed target program to the highest RHIC energy.

The latest results from RHIC have suggested the 
effect of `jet quenching' in central $Au+Au$ collisions
that acts to reduce both
the overall yield of high p$_T$ particles 
\cite{PHENIX_RAA,STAR_RAA,PHOBOS_RAA} and 
back-to-back jet correlations \cite{STAR_AAjets}. 
The presence of these dramatic effects for the most central $Au+Au$ 
collisions at $\snn = $ 130 and 200 GeV, as well as their 
absence in the cases of peripheral $Au+Au$ 
\cite{PHENIX_RAA,STAR_RAA,PHOBOS_RAA}, 
central $d+Au$ \cite{PHOBOS_RdA,STAR_dAjets}
and inclusive $d+Au$ \cite{BRAHMS_RdA,PHENIX_RdA}
have been generally well reproduced by calculations that utilize a pQCD 
framework to calculate the initial high p$_T$ production rates, coupled
with a large energy loss in the dense medium \cite{JetQuenchPaper}.
In this picture, the produced `dense medium' is responsible 
for the experimental effect, which presumably occurs only in
the large overlap volume of central $Au+Au$ collisions.

One of the intriguing overall features of the RHIC 
data, however, is that models 
solely based on parton saturation in the colliding nuclei 
describe the detailed centrality and rapidity dependence of
the measured charged particle multiplicities 
at 130 and 200 GeV \cite{PHOBOS_130_200_cent,SaturationPaper}.
If parton saturation is playing a significant role in these relativistic 
heavy ion collisions, it could also reduce the initial 
production rate of high p$_T$ particles to the extent that a
large energy loss in the dense medium is no longer 
necessary to describe the data 
for central $Au+Au$ collisions \cite{STAR_RAA,SaturationPaper}.
Since the measured total charged particle multiplicities 
are completely dominated by the emission of low p$_T$ 
($\leq$ 1.5 GeV/c) particles, one might speculate that 
the production dynamics of low p$_T$ particles is quite
different from those at high p$_T$. A study of the 
detailed centrality dependence of the bulk charged 
particle production over a large energy range may, 
therefore, provide additional constraints on models  
attempting to describe both the low and high p$_T$ 
behavior of particle production.

The PHOBOS detector configuration was the same for measurements at 
$\snn$ = 19.6 and 200 GeV.  Specifically, the detectors used in this
analysis were the centrally located Octagon barrel,
Vertex detector and the multiplane Spectrometer.  These detectors are all
constructed from silicon wafers, more details can be found in
Refs.~\cite{PHOBOS_130_200,PHOBOS_NIM}.
The primary trigger for the data reported here 
is based on $n>2$ hits in two segmented, large-area
scintillator counter arrays (Paddles) covering $3.2<|\eta|<4.5$ relative
to the nominal vertex position.  Pseudorapidity is defined as
$\eta=-\ln \tan(\theta/2)$ where $\theta$ is the polar angle to the beam
axis. This trigger was sensitive to 88\% 
of the total inelastic cross section in the 200 GeV 
data \cite{PHOBOS_130_200} and is estimated to be 80\% efficient for
19.6 GeV, from Monte Carlo (MC) studies using
Hijing \cite{HijingRef} and a full
GEANT \cite{Geant321} simulation of the PHOBOS detector.

\begin{figure} [tb] 
\resizebox{0.475\textwidth}{!}{\includegraphics{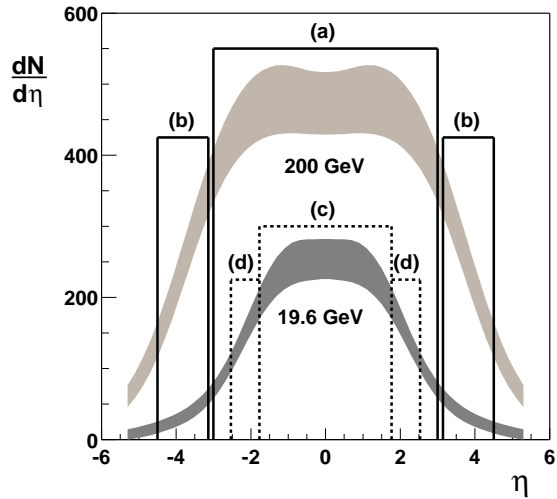}}
\caption{
Pseudorapidity density distributions from $\snn =$ 200 (light band)
and 19.6 (dark band) GeV 
$Au+Au$ collisions, for the most central 25\% of the 
cross section {\protect \cite{PHOBOS_Frag}}.  
The boxed areas (a-d) illustrate the separate regions 
in pseudorapidity used for the centrality determinations at 
each energy.  
}
\label{CentFig}
\end{figure}

The centrality determination for the PHOBOS $Au+Au$
results at $\snn = $ 130 and 200 GeV 
\cite{PHOBOS_130_200_cent,PHOBOS_130_cent}
uses the energy signals from the 
Paddle counters, which lie away from midrapidity,
as illustrated by region (b) in Fig.~\ref{CentFig}. 
These signals were found in the MC simulations 
to increase monotonically with the number of 
nucleons participating in the collision and therefore, 
through bins in the percentage of total cross section,
provide a measure of centrality.
A similar centrality measure can be created at 19.6 GeV 
by scaling the Paddle pseudorapidity range by the 
ratio of beam rapidities at 200 and 19.6 GeV, 
$y{^{Beam}_{19.6}}$/$y{^{Beam}_{200}}$ = 0.563, as 
shown in region (d) of Fig.~\ref{CentFig}.  The 
resulting $\eta$ region, $1.8 < |\eta| < 2.5$, 
lies within the Octagon detector 
coverage of $|\eta|\leq 3.2$ for collisions which occur
within $\pm$10 cm of the nominal vertex position.  
Thus, a centrality measure based on the deposited energy 
in this specific region (d)
of the Octagon was calculated for 19.6 GeV and
used in determining centrality for direct comparison to the 
original Paddle based centrality determination at 200 GeV.  

A second, independent, centrality measure was also developed for both 
the 200 and 19.6 GeV datasets.  This was
based on the energy of charged particles traversing the Octagon
within $|\eta|<3.0$ at 200 GeV, region (a) of Fig.~\ref{CentFig}, and a
`reduced' region of $|\eta|<1.8$ at 19.6 GeV, 
region (c). 

The vertex position of each event is required for the merging 
and angle correction of valid hits in the Octagon.
The primary collision vertex used was determined 
by straight-line tracks in the first six planes of the Spectrometer.
The same vertexing algorithm was used at both energies.
From MC studies of the detector, this vertex has a resolution
$\sigma_{x,y,z} \approx  0.3,0.3,0.4$~mm for central and 
$\sigma_{x,y,z} \approx  0.6,0.5,0.8$~mm for mid-peripheral 
collisions at 19.6 GeV, where z is along the beam and y is vertical.  
For more peripheral collisions, the 
efficiency falls away smoothly from 100\% with
decreasing centrality.
Additional cross-checks performed with different 
vertexing methods yielded consistent results.

Due to the requirement of a valid vertex, the resulting dataset is not only 
biased by the intrinsic trigger efficiency, but also by our (track-based) 
vertex reconstruction efficiency. 
The vertex biased detection efficiency at 200 GeV is deduced from
prior measurement \cite{PHOBOS_130_200}.  At 19.6 GeV, the efficiency
is determined from a ratio of yields (Data/MC) after 
shape matching of multiplicity distributions between
data and MC simulations in the centrality regions 
(c) or (d) from Fig.~\ref{CentFig}.  
The matching algorithm only utilizes data where
100\% efficiency in both triggering and vertexing is expected. 
The overall efficiency, including the trigger and vertexing bias,
is estimated to be $66.0\pm2.0$ \% and $55.4\pm2.0$ \% for the 
200 and 19.6 GeV data, respectively.

The final step in the centrality determination is to connect the 
experimentally deduced cross section percentiles with a well-defined
variable, such as the number of participating nucleons, $\npart$.  
In order to do this, a monotonic relation was assumed
to exist between the multiplicity distribution
in the chosen $\eta$ region and $\npart$. 
This assumption has been borne out by extensive simulations
and the experimental (inverse) correlation between multiplicity
and zero degree calorimetry.
Additionally, in order to further reduce ambiguity in 
this assumption, the measures of centrality outlined above were 
chosen specifically to lie in very different regions of pseudorapidity.  
Away from midrapidity, regions (b) and (d), particle production 
in the Hijing model depends linearly on $\npart$, as discussed in 
Ref.~\cite{Wang_Gyulassy}.  
However, the midrapidity charged particle production, 
regions (a) and (c), is not linear as predicted by 
Hijing.  Comparison of the results from these distinct regions for
centrality determination can expose any systematic effects of 
non-linearity.

\begin{figure}[tb]
\resizebox{0.475\textwidth}{!}{\includegraphics{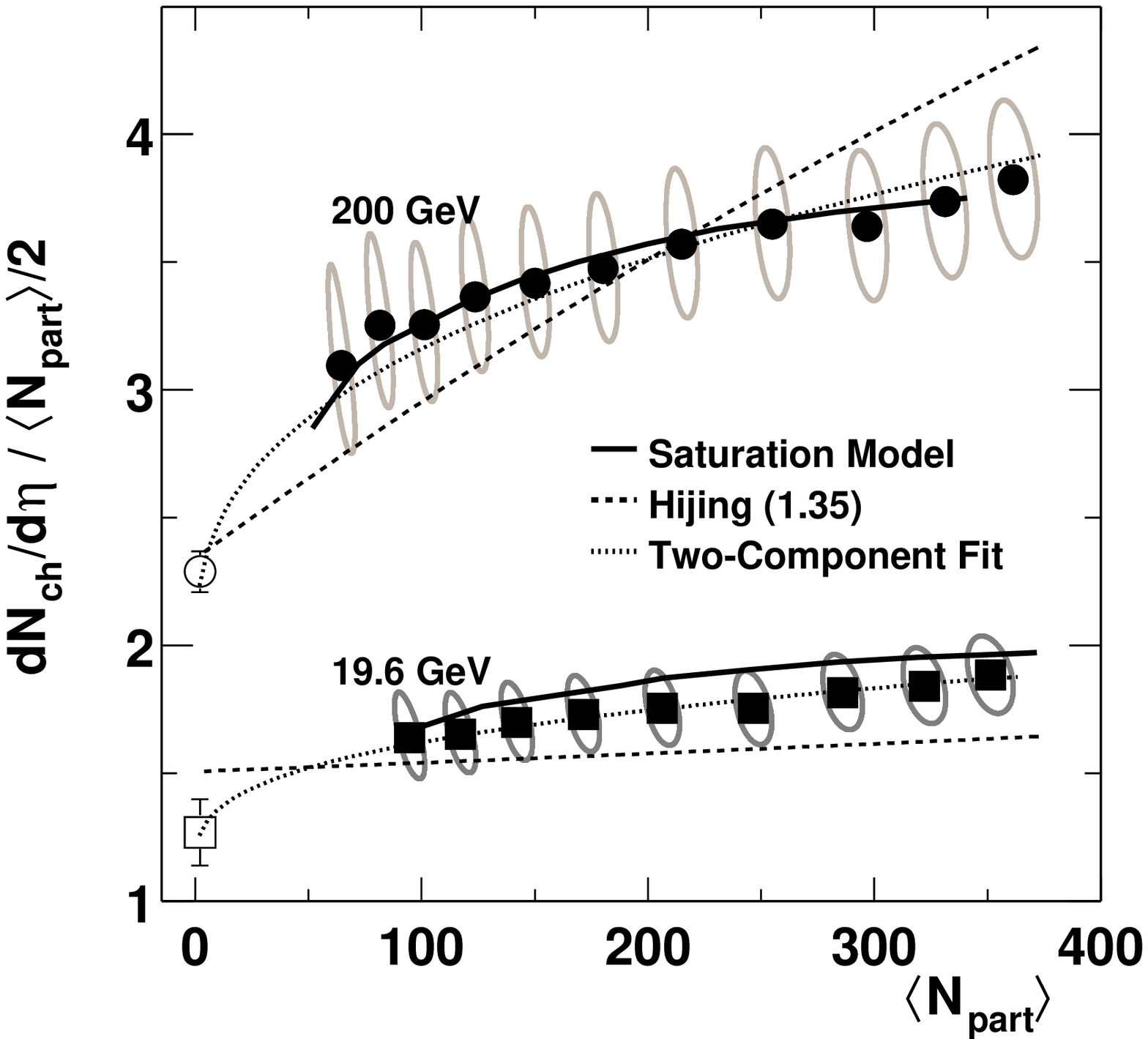}}
\caption{
The measured pseudorapidity density per participant pair
as a function of $\avenp$ 
for $Au+Au$ collisions at $\snn =$ 19.6 GeV (closed squares), 
200 GeV (closed circles).  The error ellipses around the data combine
the systematic error on $\dndetaone$ and $\avenp$.  
The open circle and square represent the $\pbarp$ results (see text). 
The three curves give two model calculations and one fit result.
As in {\protect \cite{PHOBOS_130_200_cent}}, the systematic errors represent
90\% C.L. limits. 
}
\label{figure1}
\end{figure}

The `tracklet' reconstruction method was used to obtain the 
best measure of the charged particle multiplicity at midrapidity.
A tracklet is a two-hit combination from the inner and outer Vertex 
detector layers, which points back
to the reconstructed vertex.  A Vertex detector
tracklet is only formed when the difference between the 
hits (residuals) in azimuthal angle and $\eta$ are less 
than 0.3 radians and 0.04 respectively.  The difference in the magnitude 
of these values originates from the granularity of the detector
in the respective measured directions.  The final multiplicity 
is measured in the region $|\eta|<1$ by counting all reconstructed 
tracklets and correcting for combinatorial 
background, detector acceptance and background from secondary particles
and weak decays. 
The correction factors, determined using MC simulations as 
a function of reconstructed vertex position
and number of hits in the vertex detector, were found to be
the same for each energy to within 1\%.  The results of the Vertex
detector tracklet analyses obtained using centrality regions (a) and (c) 
of Fig.~\ref{CentFig} are given in Table~\ref{Fixed_XSect_Table}.  
We obtain the same result 
for the yield per participant pair, $\dndetanp$, from centrality
regions (b) and (d).
Systematic errors for the Vertex detector tracklet results, as in
\cite{PHOBOS_130_200_cent}, are 7.5\% (90\% C.L.) at both energies. 

\begin{table*}
\caption{
\label{Fixed_XSect_Table}
Experimental results for the charged particle pseudorapidity 
density at midrapidity as a function of percentile cross section.  
The most central collisions are labeled as Bin 0-3\%.
The requirement of 100\% efficiency in both triggering and vertexing
imposes a lower limit on the reported results 
of $\langle\npart\rangle \geq$ 65 and 95 for 200 and 19.6 GeV, respectively.  
All errors represent 90\% C.L. systematic limits with the 
exception of the Ratio, for which the errors are standard
combined statistical and systematic 1-$\sigma$ uncertainties. 
}
\begin{ruledtabular}
\begin{tabular}{c|ccc|ccc|c}
&\multicolumn{3}{c|}{{\bf 200 GeV}}&
\multicolumn{3}{c|}{{\bf 19.6 GeV}}&\multicolumn{1}{c}{{\bf Ratios}} \\

Bin(\%) & $\dndeta$ & $\langle \npart \rangle$ & $\dndetanp$ & $\dndeta$
& $\langle \npart \rangle$ & $\dndetanp$ & $R_{200/19.6}$ \\ \hline 

 0 - 3 & 691 $\pm$ 52 & 361 $\pm$ 11 & 3.82 $\pm$ 0.31
       & 331 $\pm$ 24 & 351 $\pm$ 11 & 1.89 $\pm$ 0.15 & 2.03 $\pm$ 0.06\\

 3 - 6 & 619 $\pm$ 46 & 331 $\pm$ 10 & 3.74 $\pm$ 0.30
       & 297 $\pm$ 22 & 322 $\pm$ 10 & 1.84 $\pm$ 0.15 & 2.03 $\pm$ 0.06\\

 6 - 10 & 540 $\pm$ 41 & 297 $\pm$ 9 & 3.64 $\pm$ 0.30
        & 260 $\pm$ 20 & 286 $\pm$ 9 & 1.82 $\pm$ 0.15 & 2.00 $\pm$ 0.06\\

 10 - 15 & 465 $\pm$ 35 & 255 $\pm$ 8 & 3.65 $\pm$ 0.30
         & 216 $\pm$ 16 & 247 $\pm$ 8 & 1.76 $\pm$ 0.14 & 2.08 $\pm$ 0.07\\

 15 - 20 & 384 $\pm$ 29 & 215 $\pm$ 7 & 3.57 $\pm$ 0.29
         & 181 $\pm$ 14 & 206 $\pm$ 8 & 1.75 $\pm$ 0.15 & 2.04 $\pm$ 0.08\\

 20 - 25 & 313 $\pm$ 24 & 180 $\pm$ 7 & 3.47 $\pm$ 0.30
         & 148 $\pm$ 11 & 171 $\pm$ 7 & 1.73 $\pm$ 0.15 & 2.01 $\pm$ 0.09\\

 25 - 30 & 257 $\pm$ 19 & 150 $\pm$ 6 & 3.42 $\pm$ 0.29
         & 121 $\pm$ 9 & 142 $\pm$ 7 & 1.70 $\pm$ 0.15 & 2.01 $\pm$ 0.12\\

 30 - 35 & 208 $\pm$ 16 & 124 $\pm$ 6 & 3.37 $\pm$ 0.30
         & 97 $\pm$ 7 & 117 $\pm$ 7 & 1.65 $\pm$ 0.16 & 2.03 $\pm$ 0.13\\

 35 - 40 & 165 $\pm$ 12 & 101 $\pm$ 6 & 3.25 $\pm$ 0.31
         & 78 $\pm$ 6 & 95 $\pm$ 7 & 1.64 $\pm$ 0.17 & 1.98 $\pm$ 0.14\\
 
 40 - 45 & 133 $\pm$ 10 & 82 $\pm$ 6 & 3.25 $\pm$ 0.34
         & & & & \\ 

 45 - 50 & 100 $\pm$ 8 & 65 $\pm$ 6 & 3.10 $\pm$ 0.38
         & & & & \\

\end{tabular}
\end{ruledtabular}
\end{table*}

An independent analysis was also carried out at both energies
using the tracklet method applied to data obtained from 
pairs of hits in adjacent Spectrometer planes.  
This additional analysis yielded results consistent 
within 2\% across all centralities
to the results obtained using the Vertex detector.

The data for 19.6 GeV, together with the reanalyzed results 
for 200 GeV, are shown in Fig.~\ref{figure1}.  
This new result for 200 GeV has a slightly flatter dependence
on $\avenp$ than found previously \cite{PHOBOS_130_200_cent}, 
but within the quoted systematic errors.  
This flattening of the yield with $\avenp$ for central 
collisions arises entirely from the new (vertex restricted) 
centrality measures and methods detailed above.  
Also shown, at $\npart$ = 2, is the inelastic charged particle 
multiplicity obtained in $\pbarp$ collisions 
\cite{UA5_200a,UA5_200b,ISR_Data} for data measured at 200 GeV
($\dndetaone = 2.29 \pm 0.08$) and interpolated for 19.6 GeV
($\dndetaone = 1.27 \pm 0.13$). 
The $\pbarp$ data are averaged over the same pseudorapidity region,
$|\eta|<1$, as the heavy ion measurement.   Clearly, 
the yield of charged particles per participant pair
at midrapidity for the measured $Au+Au$ 
collisions is higher than found in corresponding
$\pbarp$ collisions.

The dotted line in Fig. \ref{figure1} represents a fit to the data 
using the simple two-component parameterization 
proposed in Ref. \cite{Two_Comp_Fit}: 
\[
\ensuremath{\frac{dN_{ch}}{d\eta} = n_{pp}((1-x) \frac{\avenp}{2} + x\avenc)}.
\]
\noindent
The value for $\ncoll$, the number of 
binary (nucleon-nucleon) collisions, is determined from a Glauber model
calculation and is found to depend on the number of
participants, $\npart$, through a simple power law. 
We find that $\ncoll = A \times \npart^\alpha$, with 
$A$ = 0.33 and 0.37 and $\alpha$ = 1.37 and 1.32 for 
200 GeV and 19.6 GeV respectively.
The difference in the $A$ and $\alpha$ parameters at the two energies
is due to the different measured
nucleon-nucleon cross sections at $\snn =$ 
200 GeV ($\sigma_{NN}$ = 42$\pm$1 mb) vs.
19.6 GeV ($\sigma_{NN}$ = 33$\pm$1 mb).
The remaining parameters are 
$n_{pp}$, the yield obtained in $\pbarp$ collisions, and
$x$, which represents the contribution from `hard' processes
taken to scale with $\ncoll$.

The large systematic errors on the data
preclude a simultaneous extraction of
both $n_{pp}$ and $x$ solely from the $Au+Au$ data.  
If only statistical errors are considered, the extracted parameters are 
$n_{pp}$ $\approx$ 2.7, 1.3 and $x$ $\approx$ 0.09, 0.11 at 
200 and 19.6 GeV, respectively.  

A value for $x$ can also be obtained by fixing 
$n_{pp}$ at the measured and interpolated values of 
2.29 and 1.27.  Using statistical errors,
we find $x$ = 0.145 and 0.120 for
200 and 19.6 GeV, respectively, as depicted by the dotted line in 
Fig.~\ref{figure1}. The systematic error on the fit parameter $x$ 
was determined by allowing the pp value and the data points to 
vary independently within their systematic uncertainties.  
Within the systematics, we find the fraction of hard collisions 
for both energies is consistent with a single value of
$x$ = 0.13$\pm$0.01(stat)$\pm$0.05(syst).

The equivalence of parameter $x$ at both energies, within the large errors,
is surprising as the pQCD cross section for processes with large
momentum transfers is expected to rise from $\snn =$ 19.6 to 200 GeV. 
This expectation of increasing slope in centrality with 
collision energy is shown by the dashed lines in
Fig.~\ref{figure1}, which represent Hijing predictions \cite{HijingRef}.
Although this rapid rise is not conclusively
ruled out within the systematics at $\snn =$ 200 GeV, it clearly 
does not follow the trend found
in the data.  Calculations from the parton saturation model 
\cite{SaturationPaper,Dima22GeV} (solid lines in Fig.~\ref{figure1}) 
predict a much weaker centrality dependence for 
both energies, in better agreement with the experimental data.

\begin{figure}
\resizebox{0.475\textwidth}{!}{\includegraphics{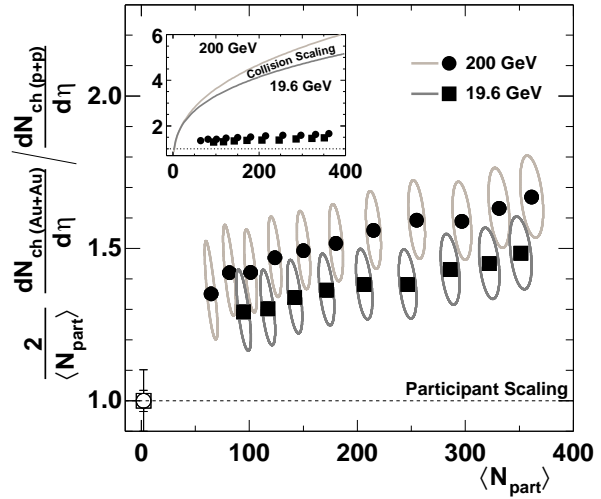}}
\caption{
Centrality dependence of the measured $Au+Au$ pseudorapidity density 
per participant pair divided by the corresponding 
value obtained in $\pbarp$ collisions.
The light and dark error ellipses on the data points 
represent the systematic errors
associated with the measurement at 200 and 19.6 GeV, respectively.  
There is an additional scale error associated with the error on 
the value of the $\pbarp$ data points of 3.5\% and 10\% 
for the 200 and 19.6 GeV data, respectively.
The figure inset has the same axes labels as the figure itself. 
The curved lines in the inset panel represent the binary collision 
($\ncoll$) scaling limit for both energies and the horizontal 
dashed line corresponds to pure $\npart$ scaling.  
As in Fig.~{\protect \ref{figure1}}, error ellipses represent
90\% C.L. limits.
}
\label{figure2}
\end{figure}

In order to gain a different perspective on the data, we scale the 
$Au+Au$ charged particle pseudorapidity density at midrapidity 
by that obtained in inelastic $\pbarp$ collisions at the same collision 
energy, shown in Fig.~\ref{figure2}.
The similarity between the two data sets is remarkable. 
The inset in Fig.~\ref{figure2} shows the data with an
expanded y-range and additional curved lines 
illustrating the expectation for 
pure binary collision ($\ncoll$) scaling at 200 and 19.6 GeV.  
The dashed horizontal line represents the expectation 
for pure participant ($\npart$) scaling.  
In this representation, it becomes clear that the
data at both energies follow a more $\npart$-like 
dependence.  

Systematic errors dominate the charged particle density
measurements at 200 and 19.6 GeV. 
We find, however, that most of these cancel in the ratio, leaving 
a baseline 3.0\% overall uncertainty. 
This occurs as both analyses were performed with exactly the 
same method, detector and with carefully matched 
centrality determinations. 
The main uncertainty comes from statistics of data and MC simulations
(2.2\%) and systematics in the primary charged 
particle detection efficiency (2.0\%).  
Smaller systematic contributions
arise from background subtraction (0.4\%)
and uncertainty in the nucleon-nucleon cross section, $\sigma_{NN}$
(0.4\%).
We also include an additional centrality dependent systematic uncertainty
that is largest in the more peripheral region and becomes
negligible for the most central. This term takes into account the 
possibility that the estimated overall efficiency error may 
not entirely cancel in the ratio.

\begin{figure}
\resizebox{0.475\textwidth}{!}{\includegraphics{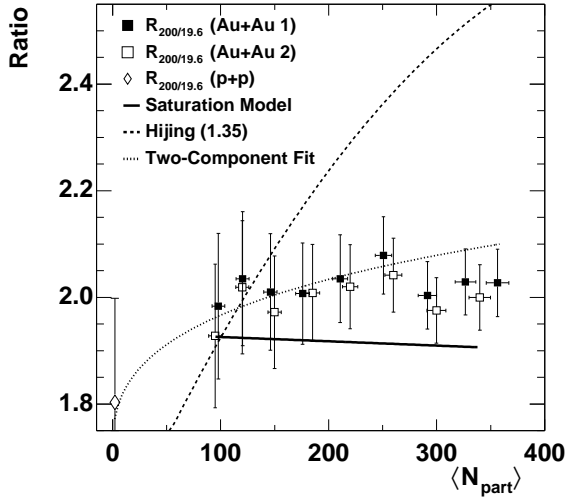}}
\caption{
The ratio, R$_{200/19.6}$, of the midrapidity
pseudorapidity density per participant pair at 200 and 19.6 GeV
versus $\avenp$.  
The data binned by fraction of cross section are shown
as closed squares and that binned by matching $\avenp$ are given as
open squares.  Also shown is the division of the inelastic
$\pbarp$ collision data (open diamond) at $\npart$ = 2. 
Curves give various calculations.
The vertical error bars are combined statistical
and systematic 1-$\sigma$ uncertainties.}
\label{figure3}
\end{figure}

As a final cross-check, the ratio of data at 200 to 
19.6 GeV, R$_{200/19.6}$ is calculated in two distinct ways. 
First, a more model-independent ratio was formed by
dividing the data at each corresponding fraction of 
total interaction cross section. This is given by the 
solid squares in Fig.~\ref{figure3} (Au+Au 1). Matching the
centrality percentile bin at each energy, however, means there will be
a difference in the deduced $\avenp$ value at 19.6 and 200 GeV
(see Table~\ref{Fixed_XSect_Table}). In this case, 
the assigned  $\avenp$ for each percentile bin given in the figure
is taken as the average of the two individual $\avenp$ values.
Second, the ratio was formed using a new set of 
centrality cuts for which each centrality bin width was varied, 
in an iterative fashion, in order to 
obtain bins at both 19.6 and 200 GeV that yield the same
calculated average $\npart$.
The data at both energies were then completely re-analyzed using 
this second set of centrality cuts.  This result is given 
by the open squares of Fig.~\ref{figure3} (Au+Au 2).

The results from the two types of ratio
calculations are shown in Fig.~\ref{figure3}, together with 
the predictions of two models and the two-component fit from 
Fig.~\ref{figure1}.
We find that both sets of data (closed and open squares) are 
in agreement, even within the significantly reduced systematic 
errors.  This level of agreement gives 
further confidence that the systematic
errors are canceling as expected.  
Additionally, we find that the slope, and hence the centrality dependence,
of both ratios is zero, within error. 
The most probable mean value of the (Au+Au 1) ratio data
is found to be \Ratio.
We remind the reader that the ratio of 200 to 
130 GeV data was found to be 
R$_{200/130}$ = 1.14$\pm$0.01(stat)$\pm$0.05(syst) 
\cite{PHOBOS_130_200_cent}.

With the reduced systematic errors on the ratio now available, 
we return to a more detailed comparison of our results
to calculations.  
As shown in Fig.~\ref{figure3}, model calculations
predict quite different centrality dependences of R$_{200/19.6}$
over the collision energy
range of 19.6 GeV to 200 GeV.
We find that the Hijing calculation gives the expected increase 
of pQCD minijet production with centrality over this energy range, 
but the predicted increase is now in strong contradiction to the data.  
The flat centrality dependence of the ratio is relatively
well described by the parton saturation model calculation.

In summary, PHOBOS has measured the charged particle pseudorapidity 
density at midrapidity ($|\eta|<1$) for $Au+Au$ collisions 
at energies of $\snn =$ 19.6 and 200 GeV.  
We find an increase in particle production per participant pair
for $Au+Au$ collisions compared to the corresponding 
inelastic $\pbarp$ values for both energies.  
The ratio of the measured yields at 200 and 19.6 GeV
shows a clear geometry scaling over the central 40\% inelastic
cross section and averages to \Ratio.
A large increase in yield from hard
processes, which contribute to multiplicity,
is not apparent in the data, even over an order of magnitude
range of collision energy.

\begin{acknowledgments}
We thank the entire RHIC accelerator staff for providing
the $Au+Au$ collisions. In particular, we express appreciation to
the BNL management for approving the one day of 19.6 GeV 
running.
This work was partially supported by U.S. DOE grants DE-AC02-98CH10886, 
DE-FG02-93ER40802, DE-FC02-94ER40818, DE-FG02-94ER40865, DE-FG02-99ER41099, 
and W-31-109-ENG-38, US NSF grants 9603486, 9722606 and 0072204, 
Polish KBN grant 2-P03B-10323, and NSC of Taiwan contract 
NSC 89-2112-M-008-024.
\end{acknowledgments}

\newcommand{\etal} {$\mathrm{\it et\ al.}$}


\begin{thebibliography}{99}

\bibitem{PHENIX_RAA}
	K.~Adcox \etal, Phys. Rev. Lett. {\bf 88} 022301 (2002)
\bibitem{STAR_RAA}
	J.~Adams \etal, Phys. Rev. Lett. {\bf 91} 172302 (2003)
\bibitem{PHOBOS_RAA}
	B.B.~Back \etal, Phys. Lett. {\bf B578} 297 (2004)	
\bibitem{STAR_AAjets}
	C.~Adler \etal, Phys. Rev. Lett. {\bf 90} 082302 (2003)
\bibitem{PHOBOS_RdA}
	B.B.~Back \etal, Phys. Rev. Lett. {\bf 91} 072302 (2003)
\bibitem{STAR_dAjets}
	J.~Adams \etal, Phys. Rev. Lett. {\bf 91} 072304 (2003)
\bibitem{BRAHMS_RdA}
	I.~Arsene \etal, Phys. Rev. Lett. {\bf 91} 072305 (2003)
\bibitem{PHENIX_RdA}
	S.S.~Adler \etal, Phys. Rev. Lett. {\bf 91} 072303 (2003)
\bibitem{JetQuenchPaper}
	M.~Gyulassy \etal, arXiv:nucl-th/0302077
\bibitem{PHOBOS_130_200_cent}
	B.B.~Back \etal, Phys. Rev. {\bf C65} 061901(R) (2002)
\bibitem{SaturationPaper}
	D.~Kharzeev and E.~Levin, Phys. Lett. {\bf B523} 79 (2001)
\bibitem{PHOBOS_130_200}
	B.B.~Back \etal, Phys. Rev. Lett. {\bf 88} 022302 (2002)
\bibitem{PHOBOS_NIM}
	B.B.~Back \etal, Nucl. Instr. and Meth. {\bf A499} 603 (2003)
\bibitem{HijingRef}
	M.~Gyulassy and X.N.~Wang, Comp. Phys. Comm. {\bf 83} 307 (1994).
	Version 1.35 is used.
\bibitem{Geant321}
	Geant~3.21, CERN Program Library, Geneva.
\bibitem{PHOBOS_130_cent}
	B.B.~Back \etal, Phys. Rev. {\bf C65} 031901(R) (2002)
\bibitem{PHOBOS_Frag}
	B.B.~Back \etal, Phys. Rev. Lett. {\bf 91} 052303 (2003)
\bibitem{Wang_Gyulassy}
	X.N.~Wang and M.~Gyulassy,
	Phys. Rev. Lett {\bf 86} 3496 (2001)
\bibitem{UA5_200a}
	UA5 Collaboration, Z. Phys {\bf C33} 1 (1986)
\bibitem{UA5_200b}
	UA5 Collaboration, Z. Phys {\bf C43} 357 (1989)
\bibitem{ISR_Data}
	W.~Thom\'{e} \etal, Nucl. Phys. {\bf B129} 365 (1977)
\bibitem{Two_Comp_Fit}
	D.~Kharzeev and M.~Nardi, Phys. Lett. {\bf B507} 121 (2001)
\bibitem{Dima22GeV}
        D.~Kharzeev, E.~Levin, M.~Nardi, arXiv:hep-ph/0111315.

\end{thebibliography}
\end{document}